\documentclass[12pt]{article}

\usepackage{amsfonts}
\usepackage{amssymb}
\usepackage{amscd}      

\def\D{\hbox{D\kern-.73em\raise.25ex\hbox{-}\raise-.25ex\hbox{ }}}
 \def\d{\hbox{d\kern-.33em\raise.75ex\hbox{-}\raise-.75ex\hbox{}}}

\def\GGG{\frak G }
\def\gr3{\GGG\,(\SSS_3)}

\def\gr2{\GGG\,(\SSS_2)}

\def\SSS{\frak S}

\def\al{{\alpha}}
\def\bet{{\beta}}

\def\gam{{\gamma}}

\def\vp{\vspace}
\def\hp{\hspace}
\def\ed{\end{document}}
\def\beq{\begin{equation}}
\def\eeq{\end{equation}}
\def\bea{\begin{eqnarray}}
\def\eea{\end{eqnarray}}
\def\ba{\begin{array}}
\def\ea{\end{array}}
\def\bi{\begin{itemize}}
\def\ei{\end{itemize}}

\def\noi{\noindent}
\def\nn{\nonumber}

 \def\qed{{\ }\hfill$\diamondsuit$}
 
\newcommand{\bp}{\noindent\begin{minipage}[c]}
\newcommand{\ep}{\end{minipage}}

\begin{document}
 \baselineskip=11pt

\title{\Large Some Aspects of Noncommutativity \\ on Real, $p$-Adic and Adelic
Spaces\footnote{Based on a talk given (by B. D.) at the Conference
'Contemporary Geometry and Related Topics', June 26 - July 2 , 2005
, Belgrade, Serbia and Montenegro}\hspace{.25mm} }
\author{{Branko Dragovich}\hspace{.25mm}\thanks{\,e-mail
address: dragovich@phy.bg.ac.yu}
\\ \normalsize{Institute of Physics, P.O. Box 57, 11001 Belgrade,}\\  \normalsize{Serbia and Montenegro}
\vspace{2mm} \\ {Zoran Raki\' c}\hspace{.25mm}\thanks{\,e-mail
address: zrakic@matf.bg.ac.yu}
\\ \normalsize{Faculty of Mathematics, P.O. Box 550, 11001 Belgrade,}\\  \normalsize{ Serbia and Montenegro}}

\date{}

\maketitle


\begin{abstract}
Classical and quantum mechanics for an extended Heisenberg algebra
with  canonical commutation relations for position and momentum
coordinates are considered. In this approach additional
noncommutativity is removed from the algebra by linear
transformation of phase space coordinates and transmitted to the
Hamiltonian (Lagrangian). This transformation does not change the
quadratic form of Hamiltonian (Lagrangian) and Feynman's path
integral maintains its well-known exact expression for quadratic
systems. The compact matrix formalism is presented and can be easily
employed in particular cases. Some $p$-adic and adelic aspects of
noncommutativity are also considered.
\end{abstract}

\bigskip

PACS numbers: 11.10.Nx, 03.65.Db, 03.65.-w

\bigskip

\section{ Introduction }

Standard $n$-dimensional quantum mechanics (QM) is based on the
Heisenberg algebra

\beq [\hat{x}_a ,\hat{p}_b ] = i\, \hbar\, \delta_{ab} , \quad
[\hat{x}_a ,\hat{x}_b ] = 0 , \quad    [\hat{p}_a ,\hat{p}_b ] = 0
,\quad a, b = 1, 2, \cdots, n , \label{1.1} \eeq

\noindent for  Hermitian operators of position $\hat{x}_a$ and
momentum  $\hat{p}_a$ coordinates  in the Hilbert space. In the
recent years there has been an intensive research of noncommutative
quantum theory with an algebra

\beq [\hat{x}_a ,\hat{p}_b ] = i\, \hbar\, \delta_{ab} , \quad
[\hat{x}_a ,\hat{x}_b ] = i\, \hbar \, \theta_{ab} , \quad
[\hat{p}_a ,\hat{p}_b ] = 0 ,\quad a, b = 1, 2, \cdots, n ,
\label{1.2} \eeq

\noindent where  $\theta_{ab}$ are constant elements of a real
antisymmetric $(\theta_{ab} =- \theta_{ab})$ $n \times n$-matrix
$\Theta$. The initial considerations of noncommutativity (NC)
(\ref{1.2}) go back to the 1930's (see, e.g. \cite{szabo1}) but the
real excitement began in the 1998 when spatial NC of the form $
[\hat{x}_a ,\hat{x}_b ] = i\, \hbar \, \theta_{ab}$ was observed in
the low energy string theory with D-branes  in a constant background
B-field (see reviews \cite{nekrasov}, \cite{szabo} and references
therein).

The NC (\ref{1.2}) leads to extended uncertainty, i.e.

\beq \Delta x_a \, \Delta p_b \geq \frac{\hbar}{2}\,\delta_{ab} ,
\quad  \Delta x_a \,  \Delta x_b \geq \frac{\hbar}{2}\,|\theta_{ab}|
, \label{1.3} \eeq

 \noindent which prevents  from simultaneous accurate measuring not only
 $x_a$ and $p_a$ but also spatial coordinates
 $x_a$ and $x_b$  $\quad (a \neq b)$. To simplify exploration
 one often takes $\theta_{ab} = \theta \,
 \varepsilon_{ab} ,$ where $(\varepsilon_{ab} ) = \mathcal E$ is the unit
 $n\times n$ antisymmetric matrix with $\varepsilon_{ab}= + 1 $ if $a < b$.
  Due to the uncertainty  $ \Delta x_a \,  \Delta x_b \geq
\frac{\hbar}{2}\,|\theta| ,  \quad (a \neq b), $ a spatial point is
not a well  defined concept and the space becomes fuzzy at distances
of the order $\sqrt{\hbar\, |\theta|} ,$ which may be much larger
than the Planck or  string length.

The most attention  in this subject has been paid to noncommutative
field theory (for reviews, see e.g. \cite{nekrasov} and
\cite{szabo}). Noncommutative quantum mechanics (NCQM) has been also
actively explored. Namely,  NCQM can be regarded as the
corresponding one-particle nonrelativistic sector of noncommutative
quantum field theory. It also provides study of  NC on simple models
and their potential experimental verification.

Models of NCQM have been mainly investigated using the Schr\"odinger
equation. The path integral method has attracted less attention,
however for systems with quadratic Lagrangians a systematic
investigation started recently ( see
\cite{dragovich1}-\cite{dragovich5} and references therein).

 We consider here $n$-dimensional NCQM which is based on the
following algebra

\bea\hp{-4mm} [ \hat{x_a},\hat{p_b}] = i\,\hbar\, ( \delta_{ab} -
\frac{1}{4}\, \theta_{ac}\, \sigma_{cb})\,, \quad
[\hat{x_a},\hat{x_b}] = i\,\hbar\, \theta_{ab}\,,\quad
[\hat{p_a},\hat{p_b}] =i\,\hbar\, \sigma_{ab}  \,, \label{1.4} \eea

\noindent  where $(\theta_{ab}) = {\Theta}$ and $(\sigma_{ab})
={\Sigma}$ are the antisymmetric matrices with constant elements.
This kind of an extended noncommutativity maintains a
 symmetry between canonical variables and yields (\ref{1.2}) in the limit
 $\sigma_{a b} \to 0$.  The
algebra (\ref{1.4})  allows  simple reduction to the usual
commutation relations

\bea\hp{-5mm} [ \hat{q_a},\hat{k_b}] = i\,\hbar\, \delta_{ab},
\qquad\quad [\hat{q_a},\hat{q_b}] = 0,\qquad\quad
[\hat{k_a},\hat{k_b}] = 0 , \label{1.5} \eea

\noindent using the following phase space linear transformation:
\bea \hat{ x_a} = \hat{q_a} - \frac{\theta_{ab}\, \hat{k_b}}{2}\,,
\qquad\qquad \hat{ p_a} = \hat{k_a} + \frac{\sigma_{ab}\,
\hat{q_b}}{2}\, , \label{1.6} \eea

\noindent where summation over repeated indices is understood. It
was shown recently \cite{dragovich6} that NC (\ref{1.4}) is suitable
to study possible dynamical control of decoherence by applying
perpendicular magnetic field to a charged particle in the plane.
This property also gives possibility to observe NC.

In this paper we present  compact general formalism of NCQM for
quadratic Lagrangians (Hamiltonians) with the (\ref{1.4}) form of
the NC. The formalism developed here is suitable for both
Schr\"odinger and Feynman approaches to quantum evolution.  There
are now many papers on some concrete models in NCQM. However, to our
best knowledge, there is no article on the evaluation of general
quadratic Lagrangians (Hamiltonians). Especially, Feynman's path
integral method to the NC has been almost ignored. Note that
quadratic Lagrangians contain an important class of physical models,
and that some of them are rather simple and exactly solvable (a free
particle, a particle in a constant field, a harmonic oscillator).
The obtained relations between coefficients in commutative and
noncommutative regimes give possibility to easily construct
effective Hamiltonians and Lagrangians in the particular
noncommutative cases.

Sec. 2 is devoted to noncommutativity on real space and contains:
matrix formalism, some expressions for quadratic Lagrangians and
Hamiltonians  as well as relations between them,  Schr\"odinger
equation and Feynman path integral. Some $p$-adic and adelic aspects
of noncommutativity are presented in Sec. 3. In the last section we
give a few concluding remarks.

\bigskip

\section{Noncommutativity on real phase space}

\noi {\bf Matrix formalism.} Let us introduce the following (matrix)
formalism. If we put $\hat{x}^T=(\hat{x}_1 ,\hat{ x}_2 , \cdots ,$ $
\cdots,\hat{x}_n)$ and $\hat{p}^T=(\hat{p}_1 , \hat{p}_2 , \cdots
,\hat{p}_n)$\,( superscript ${}^T$ denotes transposition), then all
commutators between $\hat{x}_a$ and $\hat{p}_b$ we can rewrite in
the following way: \bea\label{kom} \mbox{for}\quad \hat{P}= \left[ \ba{c} \hat{p} \\
\hat{x}
\ea \right] \quad\mbox{we put} \quad [\hat{P},\hat{P}] = \left( \ba{cc}
{[}\hat{p},\hat{p}{]} & {[}\hat{p},\hat{x}{]} \\
{[} \hat{x}, \hat{p}{]} & {[} \hat{x} , \hat{x} {]}  \ea \right) \,,
\eea where ${[}\hat{p},\hat{p}{]},\,  {[}\hat{p},\hat{x}{]},\, {[}
\hat{x}, \hat{p}{]},\, {[} \hat{x} , \hat{x} {]}$ are $\,n\times
n\,$ matrices with entries given by
$\big({[}\hat{x},\hat{p}{]}\big)_{ab}={[}\hat{x}_a,\hat{p}_b{]}\,,$
and so on. Let us mention that ${[} \hat{p}, \hat{x}{]}=- \big({[}
\hat{x}, \hat{p}{]}\big)^T\,.$ \vspace{1mm}

\noi We want to rewrite in this spirit the change of coordinates.
Namely, if we have another coordinates of the same type,
$\hat{q}^T=(\hat{q}_1 ,\hat{q}_2 , \cdots ,\hat{q}_n)$ and
$\hat{k}^T=(\hat{k}_1 , \hat{k}_2 , \cdots ,\hat{k}_n)$, and if a
linear connection is defined by \bea\label{prkor} \quad  \hat{P}=
\hat{\mathcal{A}}\, \hat{K}\,,\quad \mbox{i.e.}\quad
\left[ \ba{c} \hat{p} \\
\hat{x}
\ea \right] = \left( \ba{cc} A & B \\
C & D  \ea \right) \left[ \ba{c} \hat{k} \\
\hat{q} \ea \right]\,, \eea we want to find dependance of
$[\hat{P},\hat{P}]$ on $[\hat{K},\hat{K}]$. To do this, let us prove
the following useful statement. \vspace{3.5mm}

\noi {\bf{Lemma 1.}} Let $A$ be an arbitrary $n\times n$ matrix
whose entries commute with the coordinates of $\hat{q}^T=(\hat{q}_1
,\hat{q}_2 , \cdots ,\hat{q}_n)$ and $\hat{k}^T=(\hat{k}_1 ,
\hat{k}_2 , \cdots ,\hat{k}_n).$ Then the following commutation
relations hold \bea\nn \mathrm{(i1)}\ \
\left[A\hat{e},\hat{r}\right]=A\,\left[\hat{e},\hat{r}\right]\,,\qquad
\mathrm{(i2)} \ \
\left[\hat{e},A\hat{r}\right]=\left[\hat{e},\hat{r}\right]\,
A^T\,,\eea
 where
$\hat{e},\hat{r}\in\{ \hat{q},\hat{k}\}$.

\vspace{2.5mm} \noi \textsf{Proof.} Let us prove (i2) for
$\hat{e}=\hat{q}$ and $\hat{r}=\hat{k}.$ We have, \bea\nn \left(
\left[ \hat{q},A\hat{k}\right] \right)_{ab}
&\hspace{-2mm}=\hspace{-2mm}& \left[\hat{q}_a,A\hat{k}_b\right]=
\left[\hat{q}_a,\sum_{c=1}^n {A}_{bc}\hat{k}_c\right]=\sum_{c=1}^n
{A}_{bc}\left[\hat{q}_a, \hat{k}_c\right]= \sum_{c=1}^n
\left[\hat{q}, \hat{k}\right]_{ac}A^T_{cb}\,\\ \nn
&\hspace{-2mm}=\hspace{-2mm}& \left( \left[\hat{q}, \hat{k}\right]
A^T\right)_{ab}\,. \eea The proof of all other relations is similar.
\qed \vspace{2.5mm}

\noi Eqs. (\ref{1.6}) and (\ref{1.5}) can be rewritten in the
compact form as

\bea &&\hp{-8mm} \label{p1.6} \hat{P} = \Xi \,\, \hat{K} ,\qquad \Xi
= \left(
\begin{array}{cc}
 I & \frac12\,\, {\Sigma}  \\
- \frac12\,\, {\Theta} & I \end{array} \right) , \qquad \hat{K}=
\left(
\begin{array}{c}
 \hat{k}  \\
 \hat{q} \end{array}
\right) ,\\ \label{p1.5} && \hp{-8mm}  [\hat{K},\hat{K}]= i\,\hbar
\left(
\begin{array}{cc}
 0 & -I  \\
 I & 0 \end{array} \right)\,, \label{2.12} \eea and using Lemma 1,
relations (5), and skew-symmetricity of $\Sigma$ and $\Theta$, we
have \bea\nn [\hat{P},\hat{P}] &\hspace{-2mm}=\hspace{-2mm}& [\Xi
\,\, \hat{K},\Xi \,\, \hat{K}]=\left( \ba{cc} {[}
\hat{k}+\frac{\Sigma}{2}\,\hat{q} ,
\hat{k}+\frac{\Sigma}{2}\,\hat{q} {]}
 & {[} \hat{k}+\frac{\Sigma}{2}\,\hat{q},
\hat{q}-\frac{\Theta}{2}\,\hat{k}{]}  \\
{[}\hat{q}-\frac{\Theta}{2}\,\hat{k},\hat{k}+\frac{\Sigma}{2}\,\hat{q}{]}
& {[}\hat{q}-\frac{\Theta}{2}\,\hat{k},
\hat{q}-\frac{\Theta}{2}\,\hat{k}{]}\ea \right) \\ \label{kom1}
&\hspace{-2mm}=\hspace{-2mm}& i\,\hbar \,\left( \ba{cc}   \Sigma  &
\frac{1}{4}\,\Sigma\,\Theta -I
\\  I-\frac{1}{4}\,\Theta\,\Sigma   & \Theta
\ea \right)\,, \eea since \bea
 {[}\hat{k}+\frac{\Sigma}{2}\,\hat{q},
\hat{k}+\frac{\Sigma}{2}\,\hat{q} {]} & \hspace{-2mm}=\hspace{-2mm}
& \nn [\hat{k},\hat{k} ] + [ \frac{\Sigma}{2}\,\hat{q},\hat{k}
 ] +
 [\hat{k},\frac{\Sigma}{2}\,\hat{q}  ] +
 [\frac{\Sigma}{2}\,\hat{q},\frac{\Sigma}{2}\,\hat{q
 } ] \\
\nn &\hspace{-2mm}=\hspace{-2mm}&  \frac{\Sigma}{2}[\hat{q},\hat{k}
] +
 [\hat{k},\hat{q} ]\frac{\Sigma^T}{2}=\frac{\Sigma}{2}[\hat{q},\hat{k}
]+ [\hat{q},\hat{k} ] \frac{\Sigma}{2} = i\, \hbar\, \Sigma , \\
 {[}\hat{q}-\frac{\Theta}{2}\,\hat{k},
\hat{k}+\frac{\Sigma}{2}\,\hat{q} {]} & \hspace{-2mm}=\hspace{-2mm}
& \nn [\hat{q},\hat{k} ] - [ \frac{\Theta}{2}\,\hat{k},\hat{k}
 ] +
 [\hat{q},\frac{\Sigma}{2}\,\hat{q}  ] -
 [\frac{\Theta}{2}\,\hat{k},\frac{\Sigma}{2}\,\hat{q
 } ] \\
\nn &\hspace{-2mm}=\hspace{-2mm}&  {[}\hat{q},\hat{k} {]} -
 \frac{\Theta}{2}{[}\hat{k},\hat{q} {]}\frac{\Sigma^T}{2}
  = i\, \hbar\, \big( I-\frac14\,
 \Theta \,\Sigma\big) ,  \\ {[}\hat{q}-\frac{\Theta}{2}\,\hat{k},
\hat{q}-\frac{\Theta}{2}\,\hat{k} {]} &\hspace{-2mm}=\hspace{-2mm}&
\nn  {[}\hat{q},\hat{q} {]} - {[} \frac{\Theta}{2}\,\hat{k},\hat{q}
{]} -
 {[}\hat{q},\frac{\Theta}{2}\,\hat{k}  {]} +
 {[}\frac{\Theta}{2}\,\hat{k},\frac{\Theta}{2}\,\hat{k} {]}
  \\
\nn &\hspace{-2mm}=\hspace{-2mm}& - \frac{\Theta}{2}
[\hat{k},\hat{q} ] -
 [\hat{q},\hat{k} ]\frac{\Theta^T}{2}=
\frac{\Theta}{2} [\hat{q},\hat{k} ] +
 [\hat{q},\hat{k} ]\frac{\Theta}{2}= i\, \hbar \, \Theta
\,.
 \eea Let us note that (\ref{kom1}) contains formulas ({\ref{1.4}}) rewritten
 in the above matrix formalism.
 \vspace{3mm}

\noi {\bf Quadratic Lagrangians and Hamiltonians.} We start with
general quadratic Lagrangian for an $n$-dimensional system with
position coordinates, $x^T = (x_1 , x_2 , \cdots , x_n)$, which has
the form: \bea L(X,t) = \frac{1}{2}\, X^T \, M\, X + N^T \, X + \phi
, \label{2.2} \eea \noi where $2n\times 2n$ matrix $M$ and
$2n$-dimensional vectors $X,\, N$ are defined as \bea \hp{-7mm}
 {M } = \left(
\begin{array}{ccc}
 \alpha & \beta  \\
 \beta^T & \gamma \end{array}
\right)\, , \qquad  X^T = (\dot{x}^T\, ,\,  x^T) \, , \qquad N^T =
(\delta^T\, ,\,  \eta^T) . \label{2.3} \eea \noi where coefficients
of the $n\times n$ matrices $\alpha =((1+\delta_{ab})\,
\alpha_{ab}(t)), $ \linebreak $ \beta =(\beta_{ab}(t)),\ \gamma
=((1+\delta_{ab} )\, \gamma_{ab}(t)),$ $n$-dimensional vectors $
\delta =(\delta_{a}(t)),$ \ $\eta =(\eta_{a}(t))$ and a scalar $\phi
=\phi(t)$ are some analytic functions of  the time $t$. Matrices
$\alpha$ and $\gamma$ are symmetric,  $\alpha$ is nonsingular
$(\det\alpha \neq 0)$. \vspace{1.7mm}

\noi Using  $ p_a = {\partial L \over
\partial \dot x_a}$  one finds $ {\dot x } = {\alpha^{-1}}\, ({p} -
{\beta}\, {x } - \delta ). $ Since the function $\dot{x}$ is linear
in $p$ and $x$, then by the Legendre transformation  $ H(p,x,t)=
{p}^T\, \dot {x } - L(\dot{x},x,t)$ classical Hamiltonian is also
quadratic, i.e.

\bea H(P,t) = \frac{1}{2}\; P^T  {\mathcal M}\; P + {\mathcal N}^T
\,  P   + F , \label{2.6} \eea

\noindent where matrix ${\mathcal M}$ and vectors $P,\, {\mathcal
N}$ are

\bea \hp{-7mm}   {\mathcal M } = \left(
\begin{array}{ccc}
 A & B  \\
 B^T & C \end{array}
\right)\, , \qquad  P^T = (p^T\, ,\,  x^T) \, , \qquad {\mathcal
N}^T = (D^T\, ,\,  E^T)\, , \label{2.7} \eea

\noi and \bea \hp{-8mm} \begin{array}{ll}  {A } = {\alpha}^{-1},
\hspace{1.7cm} {B } =-\, {\alpha}^{-1}\, {\beta}, \hspace{1.7cm} {C}
=
{\beta}^T\, {\alpha}^{-1}\, {\beta} - {\gamma } , \vp{1mm} \\
  {D} =- \, {\alpha }^{-1}\,  \delta, \hspace{1.0cm}  {E } =
\beta^T \, \alpha^{-1}\,  \delta - \eta , \hspace{1.0 cm} {F} =
\displaystyle{{1\over 2}}\, {\delta}^T \, {\alpha}^{-1} {\delta}\,
- {\phi } \, .
\end{array}  \label{2.5} \eea

\noi From the symmetry of matrices $\alpha$ and $\gamma$ follows
that matrices $\,A = ((1+\delta_{ab})\, A_{ab}(t))\,$ and $\,C =
((1+\delta_{ab})\, C_{ab}(t))\,$ are also symmetric ($A^T = A ,\, \,
C^T = C$). The nonsingular $(\det {\alpha}\neq 0)$ Lagrangian
$L(\dot{x},x,t)\, $   implies nonsingular $ (\det{ A}\neq 0) $
Hamiltonian $H(p,x,t) $. Note that the inverse Legendre
transformation, i.e. $H \longrightarrow L$, is given by the same
relations (\ref{2.5}). \vspace{1.7mm}

\noi One can  show that

\bea  {\mathcal M}= \sum_{i =1}^3 \Upsilon_i^T(M)\,M\,
\Upsilon_i(M) , \label{2.8} \eea

\noi where \bea \hp{-7mm} \ba{l} \Upsilon_1(M)=\left( \ba{lr}
\al^{-1} &0\\0 & -I\ea\right), \qquad \qquad
\Upsilon_2(M)=\left( \ba{lr} 0\  &  \alpha^{-1} \beta\\
0 & 0\ea\right), \vp{4mm} \\
 \Upsilon_3(M)=\left( \ba{lr} 0\  &0\\0 & \
 i\sqrt{2}\,I\ea\right) ,\ea \label{2.9}
 \eea
 and $I$ is $n \times n$ unit matrix.
 One has also  ${\mathcal N}= Y(M) \, N,  $
 where

 \bea \hp{-7mm} Y (M)= \left(\ba{rr}
-\,\al^{-1} & 0  \\ \bet^T\,\al^{-1}& -I  \ea\right) = -\Upsilon_1
(M) + \Upsilon^T_2 (M) + i\, \sqrt{2}\,\, \Upsilon_3 (M)\,,
\label{2.10} \eea

\noi and $F = N^T\, Z(M)\, N - \phi ,$ where \bea \hp{-7mm} Z(M )
= \left(\ba{ll} \frac{1}{2}\, \al^{-1} & 0  \\ 0 & 0 \ea\right)
=\frac{1}{2}\, \Upsilon_1 (M) - \frac{i}{2\sqrt{2}}\, \Upsilon_3
(M)\, . \label{2.11}\eea

\noindent Using auxiliary matrices $\Upsilon_1 (M), \Upsilon_2 (M)$
and $\Upsilon_3 (M)$ in the above way, Hamiltonian quantities
${\mathcal M}, {\mathcal N}$ and $F$ are connected to the
corresponding Lagrangian ones $M, N$ and $\phi$. \vspace{1.7mm}

\noi Since Hamiltonians depend on canonical variables, the
transformation  (\ref{p1.6}) leads to the transformation of
Hamiltonian (\ref{2.6}). By quantization the Hamiltonian (\ref{2.6})
easily becomes

\bea H(\hat{P},t) = \frac{1}{2}\; \hat{P}^T  {\mathcal M}\; \hat{P}
+ {\mathcal N}^T \,  \hat{P}   + F , \label{q2.6} \eea

\noi because  (\ref{2.6}) is already written in the Weyl symmetric
form.

\noi Performing linear transformation (\ref{p1.6}) in (\ref{q2.6})
we again obtain quadratic quantum Hamiltonian

\bea\hp{-7mm}
 \hat
H_{\theta\sigma}(\hat{K},t) = \frac{1}{2}\,\, \hat K^T \, {\mathcal
M}_{\theta\sigma}\; \hat K + {\mathcal N}_{\theta\sigma}^T \; \hat K
+ F_{\theta\sigma} , \label{2.16} \eea

\noi where $2n\times 2n$ matrix ${\mathcal M}_{\theta\sigma}$ and
$2n$-dimensional vectors $ {\mathcal N}_{\theta\sigma}, \,
\hat{K}\,$ are

\bea\hp{-9mm}  {\mathcal M}_{\theta\sigma} = \left(
\begin{array}{ccc}
 A_{\theta\sigma} & B_{\theta\sigma}  \\
 B_{\theta\sigma}^T & C_{\theta\sigma} \end{array}
\right)   , \quad\ \ {\mathcal N}_{\theta\sigma}^T =
(D_{\theta\sigma}^T\,, \, E_{\theta\sigma}^T) , \ \ \quad \hat{K}^T
= (\hat{k}^T\,, \,  \hat{q}^T) \,, \label{2.17} \eea
 and where

\bea \hp{-8mm} \ba{ll} \displaystyle{{A}_{\theta\sigma} = {A} -
{1\over 2}\,\, {B}\, { \Theta} + {1\over 2}\,\, {\Theta}\, {B}^T
-{1\over 4}\,\, {\Theta}\, {C}\, { \Theta} ,} & \hp{5mm}
\displaystyle{{D}_{\theta\sigma} = {D} + {1\over
2}\,\, {\Theta}\, {E}}, \vp{2mm} \\
\displaystyle{{B}_{\theta\sigma} = {B} + \frac{1}{2}\, {\Theta}\,
{C} + \frac{1}{2}\, A\, \Sigma + \frac14\,\, \Theta\, B^T \,
\Sigma} , & \hp{5mm} \displaystyle{{E}_{\theta\sigma} = {E}-
\frac 12\, \Sigma\, D} ,  \vp{2mm}\\
\displaystyle{{C}_{\theta\sigma} = {C} - {1\over 2}\,\Sigma\, {B}
+ {1\over 2}\, B^T \, {\Sigma} -{1\over 4}\, {\Sigma}\, {A}\,
\Sigma}, & \hp{5mm} \displaystyle{{F}_{\theta\sigma} = {F}} .\ea
 \label{2.15}
\eea Note that for the nonsingular Hamiltonian $
H(\hat{p},\hat{x},t)$ and for sufficiently small $\theta_{ab}$ the
Hamiltonian $ H_{\theta\sigma}(\hat{k},\hat{q},t)$ is also
nonsingular. It is worth noting that $A_{\theta\sigma}$ and
$D_{\theta\sigma}$ do not depend on $\sigma$, as well as
$C_{\theta\sigma}$ and $E_{\theta\sigma}$ do not depend on $\theta$.
Classical analogue of (\ref{2.16}) maintains the same form.
\vspace{1.7mm}

 \noi From (\ref{2.6}), (\ref{p1.6}) and (\ref{2.16}) one can find connections between
${\mathcal M}_{\theta\sigma},\, {\mathcal N}_{\theta\sigma} , \, {
F}_{\theta\sigma} $ and ${\mathcal M},\, {\mathcal N}, \, F$,
which are given by the following relations:

\bea\hp{-10mm} {\mathcal M}_{\theta\sigma}=\Xi^T\,{\mathcal M}\,\,
\Xi \,, \quad\qquad {\mathcal N}_{\theta\sigma}=\Xi^T\,{\mathcal N}
, \qquad\quad {F}_{\theta\sigma} = F.  \label{2.18}
\eea\vspace{2mm}

\noi Using equations $ \dot{q}_a = \frac{\partial
H_{\theta\sigma}}{\partial k_a} $ which give $ k =
A_{\theta\sigma}^{-1}\, (\dot{q} - B_{\theta\sigma}\, q -
D_{\theta\sigma}), $ we can pass from the classical form of
Hamiltonian (\ref{2.16}) to the corresponding Lagrangian by relation
$ L_{\theta\sigma} (\dot{q},q,t) = k^T \dot{q}  - H_{\theta\sigma}
(k,q,t) . $ Note that coordinates $q_a$ and $x_a$ coincide when
$\theta =\sigma =0$. Performing necessary computations we obtain

\bea\hp{-8mm} L_{\theta\sigma}(Q,t) = \frac{1}{2}\; Q^T \,
M_{\theta\sigma}\; Q+ N_{\theta\sigma}^T \; Q + \phi_{\theta\sigma}
, \label{2.20} \eea

\noi where \bea \hp{-7mm} {M_{\theta\sigma}} = \left(
\begin{array}{ccc}
 \alpha_{\theta\sigma} & \beta_{\theta\sigma}  \\
 \beta_{\theta\sigma}^T & \gamma_{\theta\sigma} \end{array}
\right)\, ,  \quad N_{\theta\sigma}^T = (\delta_{\theta\sigma}^T\,,
\, \eta_{\theta\sigma}^T)\, , \quad Q^T = (\dot{q}^T\,, \,  q^T) \,
. \label{2.21} \eea

\noi Then the connections between ${M_{\theta\sigma}},\,
{N_{\theta\sigma}},\, \phi_{\theta\sigma} \, $ and $\, M,\,
N,\,\phi$ are given by the following relations:

\bea \hp{-9mm} \ba{l} M_{\theta\sigma} = \sum\limits_{i,j=1}^3 \,
\Xi_{ij}^T\, \, M\, \,\Xi_{ij},\qquad \qquad\hp{1.6mm}
  \Xi_{ij}=\Upsilon_i(M)\,\,\Xi\,\,\Upsilon_j(\mathcal{M}_{\theta\sigma})
, \vp{3mm} \\
 N_{\theta \sigma} = Y (\mathcal{M}_{\theta\sigma})\, \Xi^T\,
Y(M)\, N, \quad \quad \phi_{\theta \sigma} =
\mathcal{N}_{\theta\sigma}^T \, Z (\mathcal{M}_{\theta\sigma}) \,
\mathcal{N}_{\theta\sigma} - F  \, .\ea  \label{2.22} \eea

 \noi In more detail, the connection between  coefficients of the
 Lagrangians $L_{\theta\sigma}$
 and  $L$ is given by the relations:

 \bea \hp{-7mm}\ba{l} {\al}_{\theta\sigma}  = \big[\,
{\al}^{-\,1} - {\frac 12} \, (\Theta\, {\bet}^{T}\, {\al}^{-\,1}-
{\al}^{-\,1}\, {\bet}\, \Theta) -\frac14\,\Theta\,( {\bet}^{T}\,
{\al }^{-\,1}\, {\bet} - \gam)
\,\Theta \, \big]^{-\,1}\,, \vp{2mm}  \\
  {\bet}_{\theta\sigma}  =  {\al}_{\theta\sigma}\,
 \big( {\al}^{-\,1}\,
 {\bet}  - {\frac 12}\,( {\al}^{-\,1}\,\Sigma -\Theta\, {\gam}+
  \Theta\, {\bet}^{T}\, {\al}^{-\,1}\,{\bet})
  +\frac14\,\Theta\,{\bet}^{T}\,{\al}^{-\,1}\,\Sigma\big)\,, \vp{2mm} \\
 {\gam}_{\theta\sigma} = \gamma +  \bet_{\theta\sigma}^T \,
\al_{\theta\sigma}^{-\,1} \, \bet_{\theta\sigma} -  {\bet }^{T}\,
{\al}^{-\,1}\, {\bet}  + {\frac 14}\, \Sigma\,{\al}^{-\,1}\,
\Sigma\,\vp{1mm} \\ \hp{11mm}  -\ {\frac
12}\,(\Sigma\,{\al}^{-\,1}\,\bet -{\bet }^{T}\,
{\al}^{-\,1}\,\Sigma ) \,, \vp{2mm} \\
 {\delta}_{\theta\sigma}  =
{\al}_{\theta\sigma}\,\big({\al}^{-\,1}\,{\delta}+\frac
12\,\,(\Theta\,\eta- \Theta\,{\bet}^{T}\,
{\al}^{-\,1}\,{\delta}) \big)\,, \vp{2mm}  \\
 {\eta }_{\theta\sigma} = \eta + \bet_{\theta\sigma}^T \,
\al_{\theta\sigma}^{-\,1} \, \delta_{\theta\sigma} -
{\bet}^{T}\,{\al}^{-\,1}\, {\delta}  - \frac 12\,\,
\Sigma\,{\al}^{-\,1}\,{\delta}\,,\vp{2mm}  \\
 {\phi }_{\theta\sigma}  =  \phi + \frac 12\,\,
{\delta}_{\theta\sigma}^T\,  \al_{\theta\sigma}^{-\,1} \,
\delta_{\theta\sigma} - \frac 12\,\, \delta^T\,
{\al}^{-\,1}\,{\delta} \,. \ea
 \label{2.23} \eea
Note that $\alpha_{\theta \sigma}, \,  \delta_{\theta \sigma} $
and $ \phi_{\theta \sigma}$ do not depend on $\sigma$. \vp{3mm}

\bigskip

{\bf Noncommutative Schr\"odinger equation and path integral.} The
corresponding Schr\"odinger equation in this NCQM is

\bea\hp{-4mm} i\, \hbar \frac{\partial \Psi (q, t)}{\partial t} =
{{H}_{\theta\sigma} (\hat{k}, q, t)}\, \Psi (q, t) \,, \label{3.1}
\eea

\noi where $ \hat{k}_a = -\, i\, \hbar \frac{\partial}{\partial q_a}
, \, \, a = 1,2, \cdots, n$ and $ {H}_{\theta\sigma} (\hat{k}, q, t)
$ is given by (\ref{2.16}). Investigations of dynamical evolution
have been mainly performed using the Schr\"odinger equation and this
aspect of NCQM is much more developed than  the noncommutative
Feynman path integral. For this reason and importance of  the path
integral method, we will give now also a description of this
approach.

To compute a path integral, which is a basic instrument in Feynman's
approach to quantum mechanics, one can start from its Hamiltonian
formulation on the phase space. However, when Hamiltonian is a
quadratic polynomial with respect to momentum $k$ (see, e.g.
\cite{dragovich2}) such path integral on a phase space can be
reduced to the Lagrangian path integral on configuration space.
Hence, for the Hamiltonian   (\ref{2.16}) we have derived the
corresponding Lagrangian  (\ref{2.20}).

 The  standard Feynman path integral \cite{feynman} is
  \bea
\hp{-4mm} {\mathcal K} (x'',t'';x',t') =\int_{x'}^{x''} \exp \left
( \frac{i}{\hbar}\, \int_{t'}^{t''} L(\dot{q},q, t)\, dt \right
)\, {\mathcal D}q \,, \label{3.2} \eea

\noi where ${\mathcal K}(x'',t'';x',t')$ is the kernel of the
unitary evolution operator $U (t)$ and $x''=q(t''), \ x'=q(t')$
are end points.
 In ordinary quantum mechanics (OQM), Feynman's path integral for quadratic
 Lagrangians  can be
evaluated analytically and its exact  expression has the form
\cite{steiner} \bea
 \hp{-6mm} {\mathcal K}(x'',t'';x',t') =\frac{1}{(i
h)^{\frac{n}{2}}} \sqrt{\det{\left(-\frac{\partial^2 {\bar
S}}{\partial x''_a
\partial x'_b} \right)}} \exp \left(\frac{2\pi i}{h}\,{\bar
S}(x'',t'';x',t')\right), \label{3.3} \eea

\noi where $ {\bar S}(x'',t'';x',t')$ is the action for the
classical trajectory. According to (\ref{1.4}), (\ref{1.5}) and
(\ref{1.6}), NCQM related to the quantum phase space $(\hat{p} ,\,
\hat{x})$ can be regarded as an OQM on the standard phase space
$(\hat{k} ,\, \hat{q})$ and one can apply usual path integral
formalism.

A systematic path integral approach to NCQM with quadratic
Lagrangians (Hamiltonians) has been investigated during the last few
years in \cite{dragovich1}- \cite{dragovich5}. In \cite{dragovich1}
and \cite{dragovich2}, general connections between quadratic
Lagrangians and Hamiltonians for standard and $\theta \neq 0$,
$\sigma =0$ NC are established, and this formalism was applied to a
particle in the two-dimensional noncommutative plane with a constant
field and to the noncommutative harmonic oscillator. Papers
\cite{dragovich3} - \cite{dragovich5} present generalization of
articles \cite{dragovich1} and \cite{dragovich2} towards
noncommutativity (\ref{1.4}). This formalism was illustrated by  a
charged particle in a noncommutative plane with electric and
perpendicular magnetic field.

If we know Lagrangian (\ref{2.2}) and algebra (\ref{1.4}) we can
obtain the corresponding effective Lagrangian (\ref{2.20}) suitable
for the path integral in NCQM. Exploiting the Euler-Lagrange
equations \bea\nn \frac{\partial L_{\theta\sigma}}{\partial q_a}
-\frac{d}{dt} \frac{\partial L_{\theta\sigma}} {\partial{\dot q}_a}
=0 \,, \quad a = 1, 2, \cdots, n \,, \eea one can obtain the
classical trajectory $q_a =q_a (t)$ connecting end points $x' =
q(t')$ and $x''= q(t'')$, and the corresponding action is
 \bea\nn {\bar S}_{\theta\sigma}
(x'',t'';x',t') =\int_{t'}^{t''} L_{\theta\sigma} (\dot{q}, q,t)\,
dt \,.\eea Path integral in NCQM is a direct analogue of (\ref{3.3})
and its exact form expressed through quadratic action ${\bar
S_{\theta\sigma}}(x'',t'';x',t')$ is \bea {\mathcal
K}_{\theta\sigma}(x'',t'';x',t') = \frac{1}{(i h)^{\frac{n}{2}}}
\sqrt{\det{\left(-\frac{\partial^2 {\bar S_{\theta\sigma}}}{\partial
x''_a\,
\partial x'_b} \right)}}  \exp \left(\frac{2\,\pi\, i}{h}\,{\bar
S_{\theta\sigma}}(x'',t'';x',t')\right).  \label{3.4} \eea \vp{3mm}

\bigskip

\section{Noncommutativity on $p$-adic and \\ adelic spaces}

We want to explore now some possible $p$-adic and adelic
generalizations of the above noncommutativity on  real phase space.
Let us first recall some elementary properties of $p$-adic numbers
and adeles.

{\bf $p$-Adic numbers and adeles.} When we are going to consider
basic properties of $p$-adic numbers it is instructive to start with
the field $\mathbb{Q}$ of rational numbers, which is the simplest
field of numbers with characteristic $0$. $\mathbb{Q}$ also contains
all results of physical measurements. Any non-zero rational number
can be expanded into two different ways of infinite series:
\begin{equation}
\pm \, 10^n \, \sum_{k = 0}^{-\infty} a_k \, 10^k \,,  \quad  a_k
\in \{ 0, 1, 2, \cdots, 9 \}\,,          \label{4.1}
\end{equation}
\begin{equation}
p^\nu \sum_{k = 0}^{+\infty} b_k \, p^k \,,  \quad  b_k \in \{ 0, 1,
\cdots, p-1 \}\,,          \label{4.2}
\end{equation}
where $p$ is a prime number, and $n, \nu \in \mathbb{Z}$. These
expansions have the usual repetition of digits depending on rational
number but different for (\ref{4.1}) and (\ref{4.2}).

The series (\ref{4.1}) and (\ref{4.2}) are convergent with respect
to the usual absolute value $|\cdot|_\infty$ and $p$-adic norm
($p$-adic absolute value) $|\cdot|_p$ . Allowing all possibilities
for digits, as well as for integers $n $ and $\nu$, by (\ref{4.1})
and (\ref{4.2}) one can represent any real and $p$-adic number,
respectively. According to the Ostrowski theorem, the field
$\mathbb{R}$ of real numbers and the field $\mathbb{Q}_p$ of
$p$-adic numbers exhaust all possible completions of $\mathbb{Q}$.
Consequently $\mathbb{Q}$ is a dense subfield in $\mathbb{R}$ as
well as in $\mathbb{Q}_p$. These local fields have many distinct
geometric and algebraic properties. Geometry of $p$-adic numbers is
the non-Archimedean (ultrametric) one.

There are mainly two kinds of analysis on $\mathbb{Q}_p $, which are
mathematically  well developed and employed in applications. They
are related to two different mappings: $\mathbb{Q}_p \to
\mathbb{Q}_p$ and $\mathbb{Q}_p \to \mathbb{C}$. Some elementary
$p$-adic valued functions are defined by the same series as in the
real case, but the region of convergence is rather different. For
instance, $\exp_p x = \sum_{n =0}^{+\infty} \frac{x^n}{n!}$
converges in $\mathbb{Q}_p$ if $|x|_p \leq |2 p|_p $ . Derivatives
of $p$-adic valued functions are also defined as in the real case,
but using $p$-adic norm instead of the absolute value.

Very important usual complex-valued $p$-adic functions are: ({\it
i}) an additive character
\begin{equation}
\chi_p (x) = \exp 2\pi i \{ x\}_p \,,   \label{4.3}
\end{equation}
where
\begin{equation}
 \{ x\}_p = \left\{ \begin{array}{ll}   p^{-m} (a_0 + a_1\, p + \cdots +
a_{m - 1}\, p^{m-1} ) \,, \quad  & m \geq 1 \,, \\  0 \,,  \quad &
\nu \geq 0 \,,   \label{4.4}
\end{array} \right.
\end{equation}
is the fractional part of  $x$ presented in the canonical form
(\ref{4.2}); ({\it ii}) a multiplicative character
\begin{equation}
\pi_s (x) = |x|_p^s  \,, \quad  s \in \mathbb{C} \,;  \label{4.5}
\end{equation}
and  ({\it iii}) locally constant functions with compact support,
whose simple example is
\begin{equation}
\Omega (|x|_p) = \left\{  \begin{array}{ll}
                 1,   &   |x|_p \leq 1,  \\
                 0,   &   |x|_p > 1 .
                 \end{array}    \right.
                 \label{4.6}
\end{equation}

An adele $x$ is an infinite sequence
\begin{equation}
x = (x_\infty , \, x_2 , \, x_3 , \, \cdots , x_p , \,  \cdots)\,,
\label{4.7}
\end{equation}
where $x_\infty \in \mathbb{R}$ and $x_p \in \mathbb{Q}_p$ with the
restriction that for all but a finite set $\mathcal{P}$ of primes
$p$ we have $x_p \in \mathbb{Z}_p = \{ y \in \mathbb{Q}_p \, : |y|_p
\leq 1 \}$. Addition and multiplication of adeles is componentwise.
The ring of all adeles can be presented as
\begin{equation}
 {\mathbb{ A}} = \bigcup_{{\mathcal{ P}}} \mathbb{ A} (\mathcal{ P}),
 \ \ \ \  \mathbb{ A} (\mathcal{ P}) = \mathbb{ R}\times \prod_{p\in
 \mathcal{ P}} \mathbb{ Q}_p
 \times \prod_{p\not\in \mathcal{ P}} \mathbb{ Z}_p \,,         \label{4.8}
\end{equation}
where $\mathbb{Z}_p$ is the ring of $p$-adic integers. $\mathbb{ A}$
is locally compact topological space with well defined Haar measure.
There are mainly two kinds of analysis over $\mathbb{ A}$, which
generalize those on ${\mathbb R}$ and  ${\mathbb{Q}_p}$.

{\bf On $p$-adic and adelic noncommutative  analogs.}   Since 1987,
$p$-adic numbers and adeles have been successfully employed  in many
topics of modern mathematical physics (for a review, see e.g.
\cite{vladimirov1}). In particular, $p$-adic and adelic string
theory (as a review, see \cite{freund1}), quantum mechanics (see
\cite{dragovich7} as a recent review) and quantum cosmology (see
\cite{dragovich8} as a recent review) have been investigated.
 For much more information on $p$-adic numbers, adeles and
their analyses one can see \cite{schikhof1} and \cite{gelfand1}.

It is well known that combining quantum mechanics and relativity one
concludes existence of a spatial uncertainty $\Delta x$ which reads
\begin{equation}
\Delta x \geq \ell_0 = \sqrt{\frac{\hbar G}{c^3}} \sim 10^{-33} cm .
\label{4.9}
\end{equation}
The uncertainty (\ref{4.9}) may be regarded as a reason to consider
simultaneously noncommutative and $p$-adic aspects of spatial
coordinates at the Planck scale. Henceforth  we are interesting here
in p-adic analogs of the above noncommutativity considerations on
real space. Adelic approach enables to treat real and all $p$-adic
aspects of a quantum system simultaneously and as essential parts of
a more complete description. Adelic quantum mechanics was formulated
\cite{dragovich9} and successfully applied to some simple and
exactly solvable models. Here we consider also adelic approach to
noncommutativity.

 Note that instead of (\ref{1.1}) one can use an
equivalent quantization based on  relations $(h = 1)$
\begin{equation}
\chi_\infty (- \alpha_a {\hat x}_a) \, \chi_\infty (- \beta_b
\hat{p}_b) = \chi_\infty ( \alpha_a \beta_b \, \delta_{ab})\,
\chi_\infty (- \beta_b {\hat p}_b) \, \chi_\infty (- \alpha_a
\hat{x}_a), \label{4.10}
\end{equation}
\begin{equation}
\chi_\infty (- \alpha_a {\hat x}_a) \, \chi_\infty (- \alpha_b
\hat{x}_b) =  \chi_\infty (- \alpha_b {\hat x}_b) \, \chi_\infty (-
\alpha_a \hat{x}_a),    \label{4.11}
\end{equation}
\begin{equation}
\chi_\infty (- \beta_a {\hat p}_a) \, \chi_\infty (- \beta_b
\hat{p}_b) =  \chi_\infty (- \beta_b {\hat p}_b) \, \chi_\infty (-
\beta_a \hat{p}_a),    \label{4.12}
\end{equation}
where $\chi_\infty (u) =\exp (-2\pi i u)$ is real additive character
and $(\alpha, \beta)$ is a point of classical phase space.

Quantization of expressions which contain products of $x_i$ and
$p_j$ is not unique. According to the Weyl quantization  any
function $f (p,x)$, of classical canonical variables $p$ and $x$,
which has the Fourier transform $\tilde{f}(\alpha, \beta)$ becomes a
self-adjoint operator in $L_2 ({\mathbb R}^n)$ in the following way:
\begin{equation}
{\hat f} ({\hat p}, {\hat x})= \int \chi_\infty (- \alpha {\hat x} -
\beta {\hat p})  {\tilde f}(\alpha, \beta) \, d^n\alpha \, d^n\beta
. \label{4.13}
\end{equation}

It is significant that quantum mechanics on a real space can be
generalized to $p$-adic spaces for any prime  number $p$. However
there is not a unique way to perform generalization. As a result
there are two main approaches: with complex-valued and $p$-adic
valued elements of the Hilbert space on ${\mathbb Q}_p^n$. For
approach with $p$-adic valued wave functions see \cite{khrennikov1}.
$p$-Adic quantum mechanics with complex-valued wave functions  is
more suitable for connection with ordinary quantum mechanics, and in
the sequel we will refer only to this kind of $p$-adic quantum
mechanics (as a review,  see \cite{dragovich7}).

Since wave functions are complex-valued, one cannot construct a
direct analog of the Schr\"odinger equation. The Weyl approach to
quantization is suitable in $p$-adic quantum mechanics (see e.g.
\cite{dragovich7}).

Let ${\hat x}$ and ${\hat k}$ be some operators of position $x$ and
momentum $k$, respectively. Let us define operators $\chi_v (\alpha
{\hat x})$ and  $\chi_v (\beta {\hat k})$ by formulas
\begin{equation}
\chi_v (\alpha {\hat x}) \, \chi_v (a  x) = \chi_v (\alpha { x}) \,
\chi_v (a { x}) = \chi_v((a + \alpha) x) \,, \label{4.16}
\end{equation}
\begin{equation}
 \chi_v (\beta {\hat k})\,
\chi_v (b { k}) = \chi_v (\beta { k})\, \chi_v (b k) = \chi_v ((b +
\beta)) \,, \label{4.16a}
\end{equation}
where index $v$ denotes real $(v=\infty)$ and any $p$-adic case, $v
= \infty, 2,  \cdots, p, \cdots$, taking into account all
non-trivial and inequivalent valuations on ${\mathbb Q}$. It is
obvious that these operators also act on a function $\psi_v (x) \in
L_2 (\mathbb{Q}_v^n )$ , which has the Fourier transform ${\tilde
\psi}(k)$, in the following way:
\begin{equation}
\chi_v (-\alpha {\hat x})\, \psi_v (x) = \chi_v (-\alpha {\hat x})\,
\int \chi_v (- k x)\, {\tilde \psi}(k)\, d^nk = \chi_v(-\alpha x)\,
\psi_v(x) ,  \label{4.17}
\end{equation}
\begin{equation}
\chi_v (-\beta {\hat k})\, \psi_v (x) = \int \chi_v (-\beta k)\,
\chi_v (- k x)\, {\tilde \psi}(k)\, d^nk =  \psi_v(x + \beta)  ,
\label{4.18}
\end{equation}
where integration in $p$-adic case is with respect to the Haar
measure $dk$ with the properties: $d(k +a) = dk,\, \, d(ak)=|a|_p\,
dk$ and $ \int_{|k|_p\leq 1} dk =1$. Now  relations (\ref{4.10}),
(\ref{4.11}), (\ref{4.12}) can be straightforwardly generalized,
including $p$-adic cases, by replacing formally index $\infty$ by
$v$ . Thus we have
\begin{equation}
\chi_v (- \alpha_a {\hat x}_a) \, \chi_v (- \beta_b \hat{k}_b) =
\chi_v ( \alpha_a \beta_b \, \delta_{ab})\, \chi_v (- \beta_b {\hat
k}_b) \, \chi_v (- \alpha_a \hat{x}_a), \label{4.19}
\end{equation}
\begin{equation}
\chi_v (- \alpha_a {\hat x}_a) \, \chi_v (- \alpha_b \hat{x}_b) =
\chi_v (- \alpha_b {\hat x}_b) \, \chi_v (- \alpha_a \hat{x}_a),
\label{4.20}
\end{equation}
\begin{equation}
\chi_v (- \beta_a {\hat k}_a) \, \chi_v (- \beta_b \hat{k}_b) =
\chi_v (- \beta_b {\hat k}_b) \, \chi_v (- \beta_a \hat{k}_a).
\label{4.21}
\end{equation}

It is worth noting that equation (\ref{4.18}) suggests to introduce
\begin{equation}
\Big\{ \frac{\beta {\hat k}}{h} \Big\}_p^m \psi_p (x) = \int \Big\{
\frac{\beta k}{h} \Big\}_p^m \,\chi_p (-k\, x)\, {\tilde\psi}_p (k)
\, d^nk \label{4.22}
\end{equation}
which may be regarded as a new kind of the $p$-adic
pseudodifferential operator (for Vladimirov's pseudodifferential
operator, see \cite{vladimirov1}). Also equation (\ref{4.19})
suggests a $p$-adic analog of the Heisenberg algebra in the form
\begin{equation}
\Big\{\frac{\alpha_a {\hat x_a}}{h}  \Big\}_p \, \Big\{
\frac{\beta_b {\hat k_b}}{h} \Big\}_p - \Big\{ \frac{\beta_b {\hat
k_b}}{h} \Big\}_p \, \Big\{ \frac{\alpha_a {\hat x_a}}{h} \Big\}_p =
-\frac{i}{2\pi}\, \delta_{ab}\, \Big\{ \frac{\alpha_a
\beta_b}{h}\Big\}_p \,,  \label{4.23}
\end{equation}
where $h$ is the Planck constant. According to (\ref{4.23}),
$p$-adic noncommutativity depends on $\Big\{ \frac{\alpha_a
\beta_b}{h}\Big\}_p $ which is a rational number related to the size
of phase space in units of $h$. When $ \frac{\alpha_a \beta_b}{h}
\in \mathbb{Z} $ then $\Big\{ \frac{\alpha_a \beta_b}{h}\Big\}_p  =
0$ and system is $p$-adically commutative. From (\ref{4.3}) one can
derive uncertainty relation \cite{dragovich10}
\begin{equation}
\Delta \Big\{\frac{\alpha_a { x_a}}{h}  \Big\}_p \, \Delta
\Big\{\frac{\beta_b { k_b}}{h}  \Big\}_p  \geq \frac{\delta_{ab}}{4
\pi}\, \Big\{\frac{\alpha_a {\beta_b}}{h} \Big\}_p \,, \label{4.23a}
\end{equation}
which is $p$-adic analog of the first inequality in (\ref{1.3}).

Taking product of (\ref{4.19}) over all valuations $v$ we have
\begin{equation}
\prod_v \chi_v (- \alpha_a {\hat x}_a) \, \chi_v (- \beta_b
\hat{k}_b) = \prod_v \chi_v (- \beta_b {\hat k}_b) \, \chi_v (-
\alpha_a \hat{x}_a)\,,  \quad \frac{\alpha_a \, \beta_b}{h} \in
\mathbb{Q} \,, \label{4.24}
\end{equation}
since
\begin{equation}
\prod_v  \chi_v ( \alpha_a\, \beta_b\, \delta_{ab})\, =  1 \,, \quad
 \frac{\alpha_a \,
\beta_b}{h} \in \mathbb{Q} \,. \label{4.25}
\end{equation}
It follows that in an adelic quantum system with the same rational
value of $  \frac{\alpha_a \beta_b}{h}$ in real and all $p$-adic
counterparts one has commutativity between canonical operators
$\hat{x}_a$ and $\hat{k}_a$.

$p$-Adic version of (\ref{1.4}) can be obtained  from (\ref{4.19}) -
(\ref{4.21}) by adding the corresponding prefactors  on the RHS.
Adelic product  will be again commutative for rational values of
parameters $\alpha_a \,, \beta_b \,, \theta_{ab}$ and $\sigma_{ab}$.

$p$-Adic and adelic path integrals have been investigated and for
quadratic Lagrangians an analog of (\ref{3.3}) was obtained with
number field invariant form (see \cite{dragovich10a} and references
therein). For some other considerations of $p$-adic and adelic
noncommutativity including the Moyal product in the context of
scalar field theory one can see \cite{dragovich11} and
\cite{dragovich12}.

\bigskip

\section{Concluding remarks}

At the first glance one can conclude  that the phase space
transformation (\ref{1.6}) is not appropriate because it is not a
canonical one. However this transformation should not be the
canonical one since initial problem is given not only by Hamiltonian
(\ref{2.6}) but also with relations (\ref{1.4}). Using
transformations (\ref{1.6}), Hamiltonian (\ref{2.6}) with
commutation relations (\ref{1.4}) is equivalent to Hamiltonian
(\ref{2.16}) with conventional relations (\ref{1.5}).

 Let us mention  that taking $\,\sigma=0 ,\, \,  \theta=0\,$ in
the above formulas we recover expressions for the Lagrangian $\,L(X
, t),\,$ classical action $\,\bar{S} (x'',T;x',0)\, $ and
probability amplitude $\,{\mathcal K} (x'',T;x',0)\,$  of the
ordinary commutative case.

 A similar path integral approach with $\,\sigma =0\,$
 has been considered
in the context of the Aharonov-Bohm effect, the Casimir effect, a
quantum system in a rotating frame, and the Hall effect (references
on these and some other related subjects can be found in
\cite{dragovich1} - \cite{dragovich5}). Our investigation contains
all quantum-mechanical systems with quadratic Hamiltonians
(\ref{q2.6}) (Lagrangians (\ref{2.2})) on noncommutative phase space
given by relations (\ref{1.4}).

\bigskip

\section*{Acknowledgments}

The work on this article was partially supported by the Ministry of
Science and Environmental Protection, Serbia, under contract No
144032D.

\end{document}